\documentclass[conference]{IEEEtran}

\usepackage{graphicx}
\usepackage{subfig}
\usepackage{listings}
\usepackage{textcomp}
\usepackage{multirow}
\usepackage{array}
\usepackage{cite}
\usepackage{url}
\usepackage{balance}
\usepackage{amsmath}
\usepackage[pages=some]{background}
\usepackage[boxruled]{algorithm2e}

\newcolumntype{L}[1]{>{\raggedright\let\newline\\\arraybackslash\hspace{0pt}}m{#1}}
\newcolumntype{C}[1]{>{\centering\let\newline\\\arraybackslash\hspace{0pt}}m{#1}}
\newcolumntype{R}[1]{>{\raggedleft\let\newline\\\arraybackslash\hspace{0pt}}m{#1}}

\DeclareGraphicsExtensions{.pdf,.jpeg,.png,.jpg,.eps}

\begin{document}
\title{Static Graph Challenge: Subgraph Isomorphism}

\author{\IEEEauthorblockN{Siddharth Samsi,
Vijay Gadepally,
Michael Hurley,
Michael Jones,
Edward Kao,
Sanjeev Mohindra, \\
Paul Monticciolo,
Albert Reuther,
Steven Smith,
William Song,
Diane Staheli,
Jeremy Kepner \\
\IEEEauthorblockA{MIT Lincoln Laboratory, Lexington, MA}}}

\maketitle

\begin{abstract}
The rise of graph analytic systems has created a need for ways to measure and compare
the capabilities of these systems.  Graph analytics present unique scalability
difficulties.  The machine learning, high performance computing, and visual analytics
communities have wrestled with these difficulties for decades and developed
methodologies for creating challenges to move these communities forward.  The
proposed Subgraph Isomorphism Graph Challenge draws upon prior challenges from
machine learning, high performance computing, and visual analytics to create a graph
challenge that is reflective of many real-world graph analytics processing systems.
The Subgraph Isomorphism Graph Challenge is a holistic specification with multiple
integrated kernels that can be run together or independently.  Each kernel is well
defined mathematically and can be implemented in any programming environment.
Subgraph isomorphism  is amenable to both  vertex-centric implementations and
array-based implementations (e.g., using the GraphBLAS.org standard).  The
computations are simple enough that performance predictions can be made based on
simple computing hardware models.  The surrounding kernels provide the context for
each kernel that allows rigorous definition of both the input and the output for each
kernel.  Furthermore, since the proposed graph challenge is scalable in both problem
size and hardware, it can be used to measure and quantitatively compare a wide range
of present day and future systems.  Serial implementations in C++, Python, Python
with Pandas, Matlab, Octave, and Julia have been implemented and their single
threaded performance have been measured.  Specifications, data, and software are
publicly available at GraphChallenge.org.
\end{abstract}


\IEEEpeerreviewmaketitle

\section{Introduction}
\let\thefootnote\relax\footnotetext{This material is based upon work supported  by
  the Defense Advanced Research Projects Agency under Air Force Contract
  No. FA8721-05-C-0002. Any opinions, findings and conclusions or recommendations
  expressed in this material are those of the authors and do not necessarily
  reflect the views of the Department of Defense.}
Increasingly, large amounts of data are collected from social media, sensor feeds
(e.g. cameras), and scientific instruments and are being analyzed with graph
analytics to reveal the complex relationships between different data feeds
\cite{darpahive}. Many graph analytics are executed in large data centers on large
cached or static data sets. The processing required is a function of both the size of
the graph and the type of data being processed. There is also an increasing need to make decisions
in real-time to understanding how relationships represented in a graph
evolve. Previous research on streaming graph analytics has been limited by
the amount of processing required. Graph analytic updates must be performed at the
speed of the incoming data. The sparseness of graph data can make the application of
graph analytics on current processors extremely inefficient. This inefficiency has
either limited the size of the graph that can be addressed to only what can be held in main memory or
requires an extremely large cluster of computers to make up for this inefficiency.
The development of a novel graph analytics system has the potential to enable the
discovery of relationships as they unfold in the field rather than relying on
forensic analysis in data centers. Furthermore, data scientists can explore
associations previously thought impractical due to the amount of processing required.
The Subgraph Isomorphism Graph Challenge and the Stochastic Block Partition Challenge~\cite{ed} (see http://GraphChallenge.org) seek to enable a
new generation of graph analysis systems by highlighting the benefits of novel
innovations in these systems.

Challenges such as YOHO~\cite{yoho}, MNIST~\cite{mnist}, HPC
Challenge~\cite{hpcc}, ImageNet~\cite{imagenet} and VAST~\cite{vast1,vast2} have
played important roles in driving progress in fields as diverse as machine
learning, high performance computing and visual analytics. YOHO is the Linguistic Data
Consortium database for voice verification systems and has been a critical enabler of speech research.
The MNIST database of handwritten letters has been a bedrock of the computer vision
research community for two decades. HPC Challenge has been used by the supercomputing
community to benchmark and acceptance test the largest systems in the world as well as
stimulate research on the new parallel programing environments. ImageNet populated an image dataset according to the WordNet hierarchy consisting of over 100,000
meaningful concepts (called synonym sets or synsets)~\cite{imagenet} with an average of 1000 images per synset and has become a critical enabler of vision research. The VAST Challenge is an annual
visual analytics challenge that has been held every year since 2006; each year, VAST offers a
new topic and submissions are processed like conference papers.  The
Subgraph Isomorphism Graph Challenge seeks to draw on the best of these challenges,
but particularly the VAST Challenge in order to highlight innovations across the
algorithms, software, hardware, and systems spectrum.

The focus on graph analytics allows the Subgraph Isomorphism Graph Challenge to also
draw upon significant work from the graph benchmarking community.  The Graph500
(Graph500.org) benchmark (based on \cite{bader2006designing}) provides a scalable
power-law graph generator \cite{leskovec2005realistic} (used to build the world's largest synthetic graphs) with the goal of optimizing the rate of building a tree of the graph.  The
Firehose benchmark (see http://firehose.sandia.gov) simulates computer network traffic for
performing real-time analytics on network traffic.  The PageRank Pipeline benchmark
\cite{dreher2016pagerank,bisson2016cuda} uses the Graph500 generator (or any other
graph) and provides reference implementations in multiple programming languages to allow users to
optimize the rate of computing PageRank (1st eigenvector) on a graph. Finally,
miniTri (see mantevo.org) \cite{wolf2015,wolf2016} takes an arbitrary graph as input
and optimizes the time to count triangles.

The organization of the rest of this paper is as follows.  Section II describes
examples of data sets that are relevant to the Graph Challenge.  Section III provides
details on the specifics of the subgraph isomorphism problem.  Section IV gives
example algorithms and implementations that can be used to solve the specific
subgraph isomorphism problem.  Section V presents metrics and preliminary serial
performance results of the example implementation over a range of graphs.  Section VI
summarize the work and describes future directions.

\section{Data Sets}
\label{sec:datasets}
Scale is an important driver of the Graph Challenge and graphs with billions to
trillions of edges are of keen interest.  The Graph Challenge is designed to work on
arbitrary graphs drawn from both real-world data sets as well as simulated data sets.
Examples of real-world data sets include the Stanford Large Network Dataset
Collection (see http://snap.stanford.edu/data), the AWS Public Data Sets (see
aws.amazon.com/public-data-sets), and the Yahoo! Webscope Datasets (see
webscope.sandbox.yahoo.com).  These real-world data sets cover a wide range of
applications and data sizes.  While real-world data sets have many contextual
benefits, synthetic data sets allow the largest possible graphs to be readily
generated. Examples of synthetic data sets include Graph500, Block Two-level
Erdos-Renyi graph model (BTER) \cite{seshadhri2012community}, Pure Kronecker Graphs
\cite{kepner2011graph}, and  Perfect Power Law graphs
\cite{kepner2012perfect,gadepally2015using}.

The focus of the Graph Challenge is on graph analytics.  While parsing and formatting
complex graph data is necessary in any graph analysis system, these data sets are
made available to the community in a variety of pre-parsed formats to minimize the
amount of parsing and formatting  required by Graph Challenge particpants.  The public data
are available in a variety of formats such as linked list, tab separated, and
labeled/unlabeled. The Graph Challenge will provide data in two formats: tab separated value
(TSV) triples in an ASCII file and MMIO ASCII format (see
math.nist.gov/MatrixMarket)~\cite{boisvert1996matrix}.  In each case, the data have
been parsed so that all vertices are integers from 1 to the total number of vertices
$n$ in the graphs and all edges weights are set to a value of 1.  In addition to this, all self loops were removed from the original datasets. For directed graphs, additional edges in the opposite direction were added to make the graphs un-directed. The edges are stored in TSV files as triples of tab separated numeric
strings with a newline between each edge. For example, let all the starting and ending vertices be stored in the $m$ element vectors $\bf{u}$ and $\bf{v}$. The edges of the graph are stored in the TSV file as shown here:

\begin{eqnarray*}
 {\bf u}(1) & {\bf v}(1) & 1 \\
  \vdots & \vdots & \vdots \\
 {\bf u}(m) & {\bf v}(m) & 1 \\
\end{eqnarray*}
where $i = 1,\ldots,m$, ${\bf u}(i) \in \{1,\ldots,n\}$, and ${\bf v}(i) \in \{1,\ldots,n\}$.

Filtering on edge/vertex labels is often used when available to reduce the search
space of many graph analytics.  Filtering can be applied at initialization, during
intermediate steps, or at the vertex level. Such filtering can be invaluable and is
highly problem specific.
Some Graph Challenge data sets in their original forms have labels and some data sets
are unlabeled.  Some users will want to filter on labels and these innovations are
encouraged.  The provided example implementations will all work without labels.

\section{Static Graph Isomorphism Challenge}
\label{sec:static_challenge}

\subsection{Triangle counting}
\label{sec:tri}
\begin{figure}[ht]
\centering
\includegraphics[height=3.5pc]{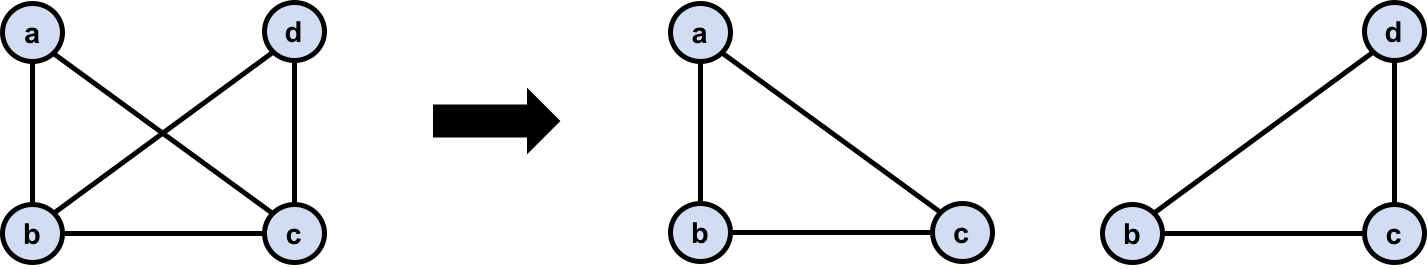}
\caption{The graph shown in this example contains two triangles consisting of nodes \{a,b,c\} and \{b,c,d\}.}
\label{fig:triangle}
\end{figure}

Triangles are the most basic, trivial sub-graph. A triangle can be defined as a
set of three mutually adjacent vertices in a graph. As shown in Figure~\ref{fig:triangle}, the graph \textbf{G} contains two triangles comprising of nodes \{a,b,c\} and \{b,c,d\}. The number of triangles in a graph
is an important metric used in applications such social network mining, link
classification and recommendation, cyber security, functional biology and spam
detection ~\cite{pavan2013}.

The number of triangles in a given graph \textbf{G} can be calculated in several
ways. We highlight two algorithms based on linear algebra primitives. The first
algorithm proposed by Wolf, et. al.~\cite{wolf2015} uses an overloaded matrix
multiplication approach on the adjacency and incidence matrices of the graph and is
shown in Listing ~\ref{triangle_1}. The second approach proposed by Burkhardt,
et. al.~\cite{burkhardt2016} uses only the adjacency matrix of the given graph and is
shown in Listing ~\ref{triangle_2}.

Another algorithm for triangle counting based on a masked matrix multiplication
approach has been proposed by Azad et al~\cite{gilbert2015}. The serial version of
this algorithm based on the MapReduce implementation by Cohen et al~\cite{cohen2009}
is shown in  Listing ~\ref{triangle_3}. Finally, a comparison of triangle counting
algorithms based on subgraph matching, programmable graph analytics and a  matrix
formulation based on sparse matrix multiplication can be found in~\cite{wang2016tc}.

\begin{algorithm}[h]
  \KwData{Adjacency matrix \textbf{A} and incidence matrix \textbf{E}}
  \KwResult{Number of triangles in graph \textbf{G}}
  initialization\;
  $\textbf{C} = \textbf{A}\textbf{E}$\\
  $n_{T} = nnz(C)/3$\\
  \vspace{.25cm}
  Multiplication is overloaded such that \\
  $\textbf{C}(i,j) = \{i, x, y\}$ iff \\
  $\textbf{A}(i,x) = \textbf{A}(i,y) = 1$ \& $\textbf{E}(x,j) = \textbf{E}(y,j) = 1$ \\
  \vspace{.2cm}
  \caption{Array based implementation of triangle counting algorithm using the adjacency and incidence matrix of a graph~\cite{wolf2015}.}
  \label{triangle_1}
\end{algorithm}

\begin{algorithm}[ht]
\KwData{Adjacency matrix \textbf{A}}
\KwResult{Number of triangles in graph \textbf{G}}
initialization\;
$\textbf{C} = \textbf{A}^2 \circ \textbf{A}$ \\
$n_{T} = \sum_{ij}^{ } (\textbf{C}) / 6$ \\
\vspace{1em}
Here, $\circ$ denotes element-wise multiplication\\
\vspace{.2cm}
\caption{Array based implementation of triangle counting algorithm using only the adjacency matrix of a graph~\cite{burkhardt2016}.}
\label{triangle_2}
\end{algorithm}

The number of triangles in a given graph \textbf{G} can be calculated in several
ways. We highlight two algorithms based on linear algebra primitives. The first
algorithm proposed by Wolf, et. al.~\cite{wolf2015} uses an overloaded matrix
multiplication approach on the adjacency and incidence matrices of the graph and is
shown in Algorithm~\ref{triangle_1}. The second approach proposed by Burkhardt,
et. al.~\cite{burkhardt2016} uses only the adjacency matrix of the given graph and is
shown in Algorithm~\ref{triangle_2}.

Another algorithm for triangle counting based on a masked matrix multiplication
approach has been proposed by Azad, et. al.~\cite{gilbert2015}. The serial version of
this algorithm based on the MapReduce implementation by Cohen, et. al.~\cite{cohen2009}
is shown in  Algorithm~\ref{triangle_3}. Finally, a comparison of triangle counting
algorithms based on subgraph matching, programmable graph analytics and a matrix
formulation based on sparse matrix multiplication can be found in~\cite{wang2016tc}.

\begin{algorithm}[h]
\KwData{Adjacency matrix \textbf{A}}
\KwResult{Number of triangles in graph \textbf{G}}
initialization\;
$(\textbf{L}, \textbf{U}) \leftarrow \textbf{A}$\\
$\textbf{B} = \textbf{L}\textbf{U} $ \\
$\textbf{C} = \textbf{A} \circ \textbf{B} $ \\
$n_{T} = \sum_{ij}^{ } (\textbf{C}) / 2$ \\
\vspace{1em}
Here, $\circ$ denotes element-wise multiplication\\
\vspace{.2cm}
\caption{Serial version of triangle counting algorithm based on MapReduce version by Cohen, et. al.~\cite{cohen2009} and ~\cite{gilbert2015}.}
\label{triangle_3}
\end{algorithm}

\subsection{k-Truss}
\label{sec:ktruss}
\begin{figure}[ht]
\centering
\includegraphics[height=7pc]{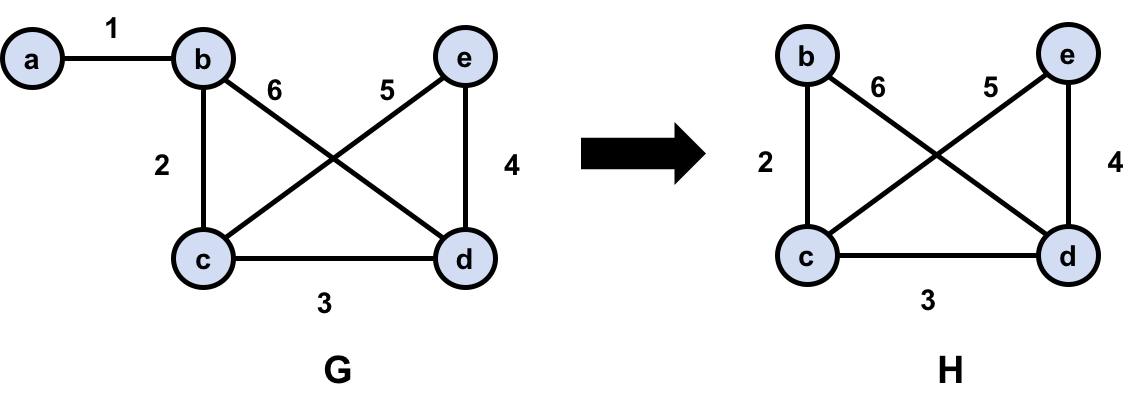}
\caption{Graph \textbf{G} and its 3-truss: Each edge in Graph \textbf{H} is part of at-least one triangle, where k = 3.}
\label{fig:ktruss}
\end{figure}

Given a graph $\textbf{G}$, a k-truss is a subgraph such that each edge is contained in at least $(k-2)$ triangles in the same
subgraph~\cite{cohen2008trusses,wang2012}. Figure~\ref{fig:ktruss} shows graph \textbf{G} and the 3-truss of this graph represented by the graph \textbf{H}. As seen in this figure, each edge of \textbf{H} is part of only one triangle. The edge labelled \textit{1} violates the k-truss condition and is removed.

\begin{algorithm}[ht]
\KwData{Unoriented incidence matrix \textbf{E} and integer $k$}
\KwResult{Incidence matrix of k-truss subgraph $E_k$}
initialization\;
$d = sum(\textbf{E})$\\
$\textbf{A} = \textbf{E}^{T}\textbf{E} - diag(d)$\\
$\textbf{R} = \textbf{E}\textbf{A}$\\
$s = (\textbf{R} == 2)\textbf{1}$\\
$x = find(s < k-2 )$\\
\While{x is not empty}{
  $\textbf{E}_x = \textbf{E}(x, :)$\\
  $\textbf{E} = \textbf{E}(x_c, :)$\\
  $d_x = sum(\textbf{E}_x)$\\
  $\textbf{R} = \textbf{R}(x_c, :)$\\
  $\textbf{R} = \textbf{R} - \textbf{E}[\textbf{E}_x^{T}\textbf{E}_x - diag(d_x)]$\\
  $s = (\textbf{R}==2)\textbf{1}$\\
  $x = find(s < k-2)$
}
\caption{Array based implementation of $k$-Truss algorithm.}
\label{ktruss_algorithm}
\end{algorithm}

Computing the truss decomposition of a graph involves finding the maximal ${k}$-truss for all ${k \ge 2}$~\cite{graphulo} and is summarized in Algorithm~\ref{ktruss_algorithm}. Let \textbf{E} be the unoriented incidence matrix and \textbf{A} be the
adjacency matrix of graph \textbf{G}. Each row of \textbf{E} has a 1 in the
column of each associated vertex. To get the support for this edge, we need the
overlap of the neighborhoods of these vertices. If the rows of the adjacency
matrix \textbf{A} associated with the two vertices are summed, this corresponds
to the entries that are equal to 2. Summing these rows is equivalent to
multiplying \textbf{A} on the left by the edge’s row in \textbf{E}. Therefore, to get the support
for each edge, we can compute \textbf{EA}, apply to each entry a function that maps 2 to
1 and all other values to 0, and sum each row of the resulting matrix. Note also
that

\begin{equation*}
\textbf{A} = \textbf{E}^{T}\textbf{E} - diag(\textbf{E}^{T}\textbf{E})
\end{equation*}

\noindent which allows us to recompute \textbf{EA} after edge removal without performing
the full matrix multiplication. Pseudocode for the array based implementation of
the k-truss algorithm is shown in Algorithm~\ref{ktruss_algorithm}. Within the pseudocode,
$x_c$ refers to the complement of $x$ in the set of row indices. Semantics of the MATLAB functions \texttt{find}, \texttt{sum} and \texttt{diag} can be found in the MATLAB documentation~\cite{matlabdocs}.
This algorithm can return the full truss decomposition by computing the truss with $k = 3$ on
the full graph, then passing the resulting incidence matrix to the algorithm
with an incremented $k$. This procedure will continue until the resulting
incidence matrix is empty.

\section{Computational Metrics}
\label{sec:metrics}
Submissions to the static graph isomorphism challenge will be evaluated on the
basis of two metrics: Correctness and Performance.

\subsection{Correctness}
\label{sec:correctness}
For the triangle counting kernel, correctness is evaluated by comparing the
reported triangle count with the ground truth since the number of triangles for
a given graph is exact. Similarly, for the k-truss kernel, correctness is based
on enumerating all k-trusses for a given graph and comparing with the exact
enumeration for said graph.

\subsection{Performance}
\label{sec:perf}
The performance of the algorithm implementation should be reported in terms of the
following metrics:
\begin{itemize}
\item Total number of edges in the given graph: This measures the amount of data processed
\item Execution time: Total time required to count the triangles or compute the
  k-truss of the given graph. Time required for reading graph data from a file is not
  included in this time.
\item Rate: Measures the throughput of the implementation as the ratio of the number of
  edges in the graph to the execution time.
\item Energy: Total amount of energy consumption in watts for the computation.
\item Rate per energy (edges/second/Watt): Measures the throughput achieved per unit
  of energy consumed.
\item Memory: Specifies the amount of memory required for the computation.
\item Processor: Number and type of processors used in the computation.
\end{itemize}

\subsection{Preliminary Benchmarking Results}
Serial implementations of the Graph Challenge benchmarks were developed and
tested on the MIT SuperCloud and the Lincoln Laboratory TX-Green
supercomputer. Serial versions of the two graph benchmarks (triangle counting and
k-truss) were implemented in MATLAB, python and Julia. Both benchmarks described in
the earlier section were tested on Intel Xeon E5-2683 based compute nodes with 256 GB of
RAM. The compute nodes were scheduled for exclusive access so that the benchmark
process had exclusive access to all hardware resources. Table~\ref{table:sloc} shows
the source lines of code for the k-truss and triangle counting benchmarks.

\begin{table}[t]
\centering
\def\arraystretch{1.25}
\subfloat[Triangle counting]{
  \begin{tabular}{|c|c|}
    \hline
    \textbf{Language} & \textbf{Count} \\
    \hline
    MATLAB & 38\\
    \hline
    Octave & 38\\
    \hline
    python & 55\\
    \hline
    Julia  & 34\\
    \hline
  \end{tabular}
}
\qquad
\subfloat[k-Truss]{
    \begin{tabular}{|c|c|}
      \hline
      \textbf{Language} & \textbf{Count} \\
      \hline
      MATLAB & 40\\
      \hline
      Octave & 40\\
      \hline
      python & 110\\
      \hline
      Julia  & 45\\
      \hline
    \end{tabular}
}
\caption{Source lines of code~\cite{sclc} for implementations of triangle counting and k-truss
  algorithms in MATLAB, GNU Octave, python and Julia.}
\label{table:sloc}
\end{table}

\begin{table}
\centering
\def\arraystretch{1.25}
\begin{tabular}{|c|c|c|c|}
\hline
\textbf{n} & \textbf{Node count (M)} & \textbf{Edge Count} & \textbf{Triangles} \\
\hline
     8   &        65536  &         260610    &       520200 \\
\hline
     9   &       262144  &        1045506    &      2088968 \\
\hline
    10   &      1048576  &        4188162    &      8372232 \\
\hline
    11   &      4194304  &       16764930    &     33521672 \\
\hline
    12   &     16777216  &       67084290    &    134152200 \\
\hline
    13   &     67108864  &      268386306    &    536739848  \\
\hline
\end{tabular}
\caption{Node, edge and triangle counts for synthetic graphs used for testing initial implementations of the triangle counting and k-truss algorithms.}
\label{table:image_graphs}
\end{table}

\subsubsection{Benchmarking on Synthetic Graph Data}
\label{sec:synthetic_data}
\begin{figure}[h]
  \centering
  \subfloat[Synthetic graph with 4 nodes]{
    \includegraphics[height=7pc]{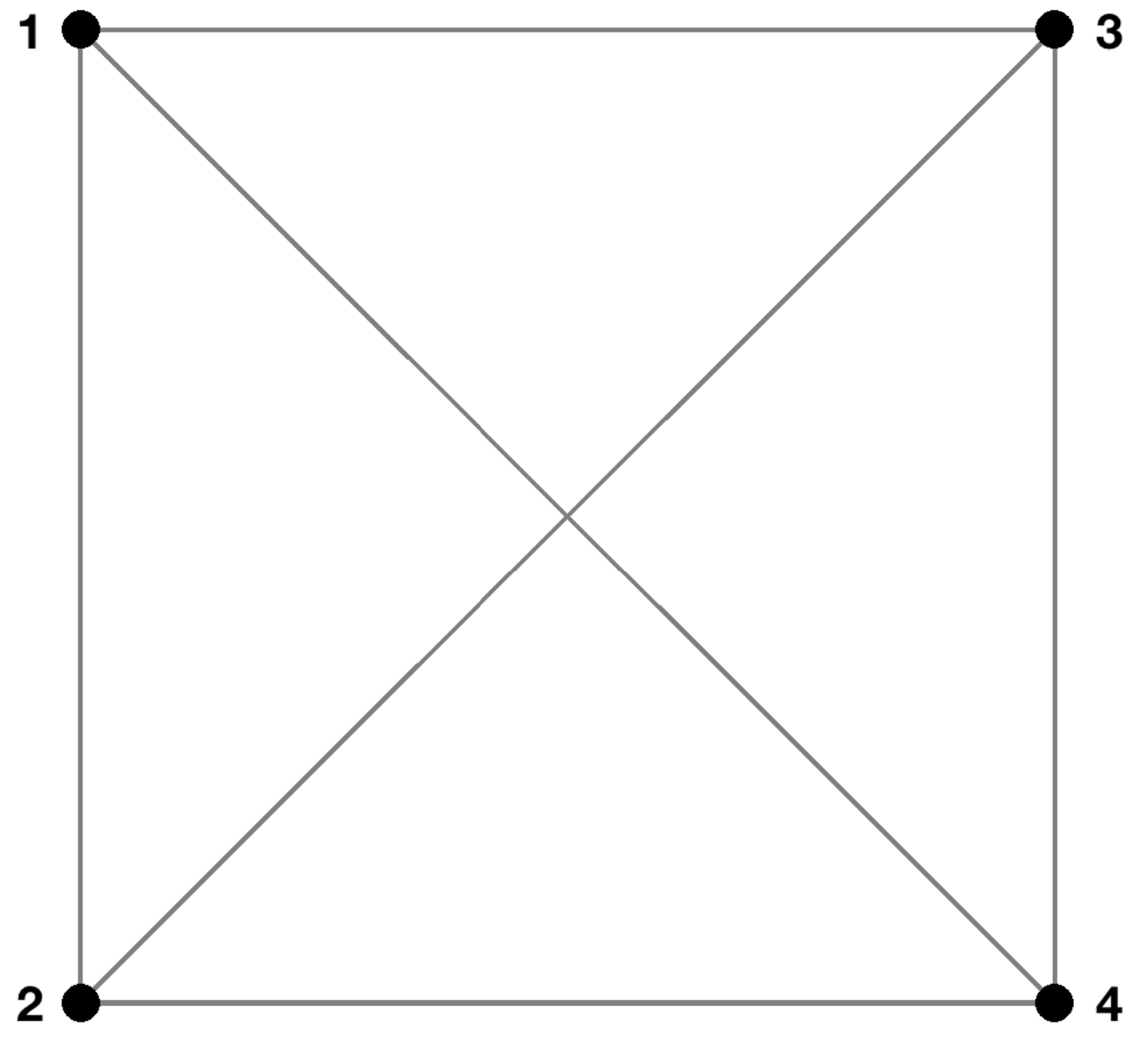}
  }
\qquad
  \subfloat[Synthetic graph with 64 nodes]{
    \includegraphics[height=7pc]{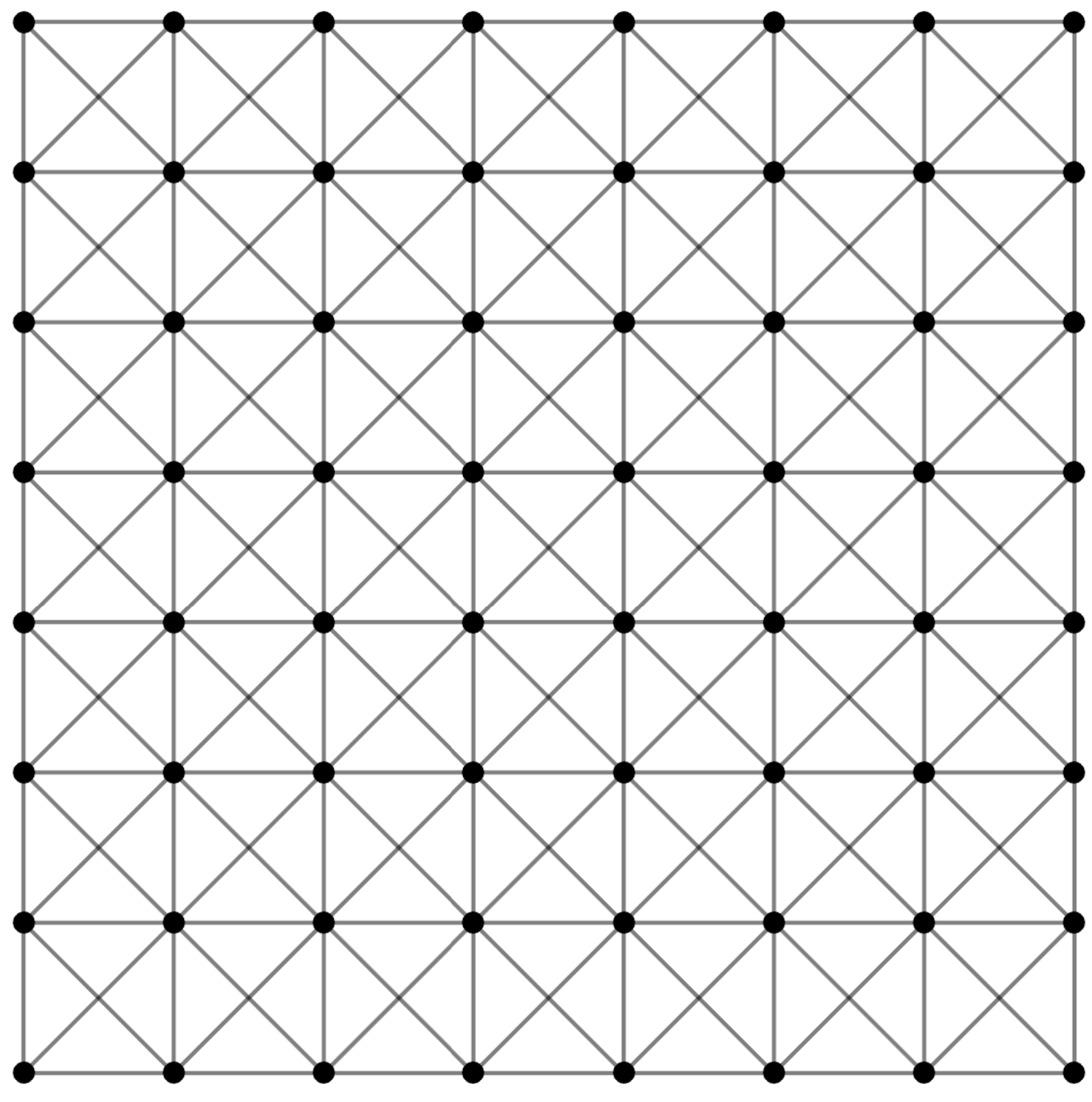}
    \label{subfig:8x8}
  }
\caption{Examples of synthetic graphs: Graphs are generated using $MxM$ images with each pixel in the image being a node in the graph. Each pixel is connected to its 8 neighbors by an undirected edge. Pixels on the boundary only have 3 neighbors.}
\label{fig:synthetic_graphs}
\end{figure}

\begin{figure}[ht]
\centering
\includegraphics[height=10pc]{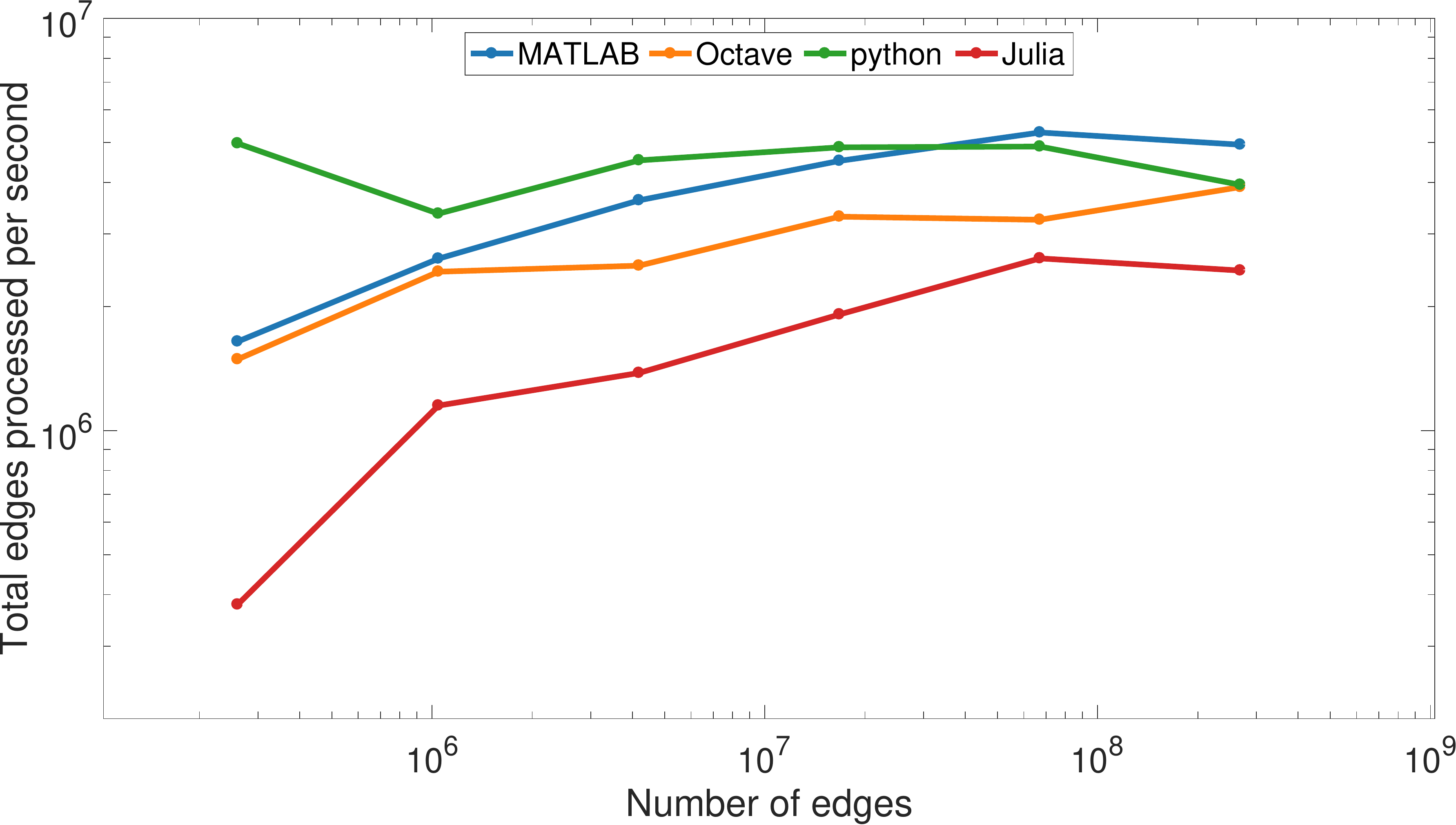}
\caption{Triangle-counting single core performance on synthetic data.}
\label{fig:triangle_synthetic_edgesort}
\end{figure}

\begin{figure}[ht]
\centering
\includegraphics[height=10pc]{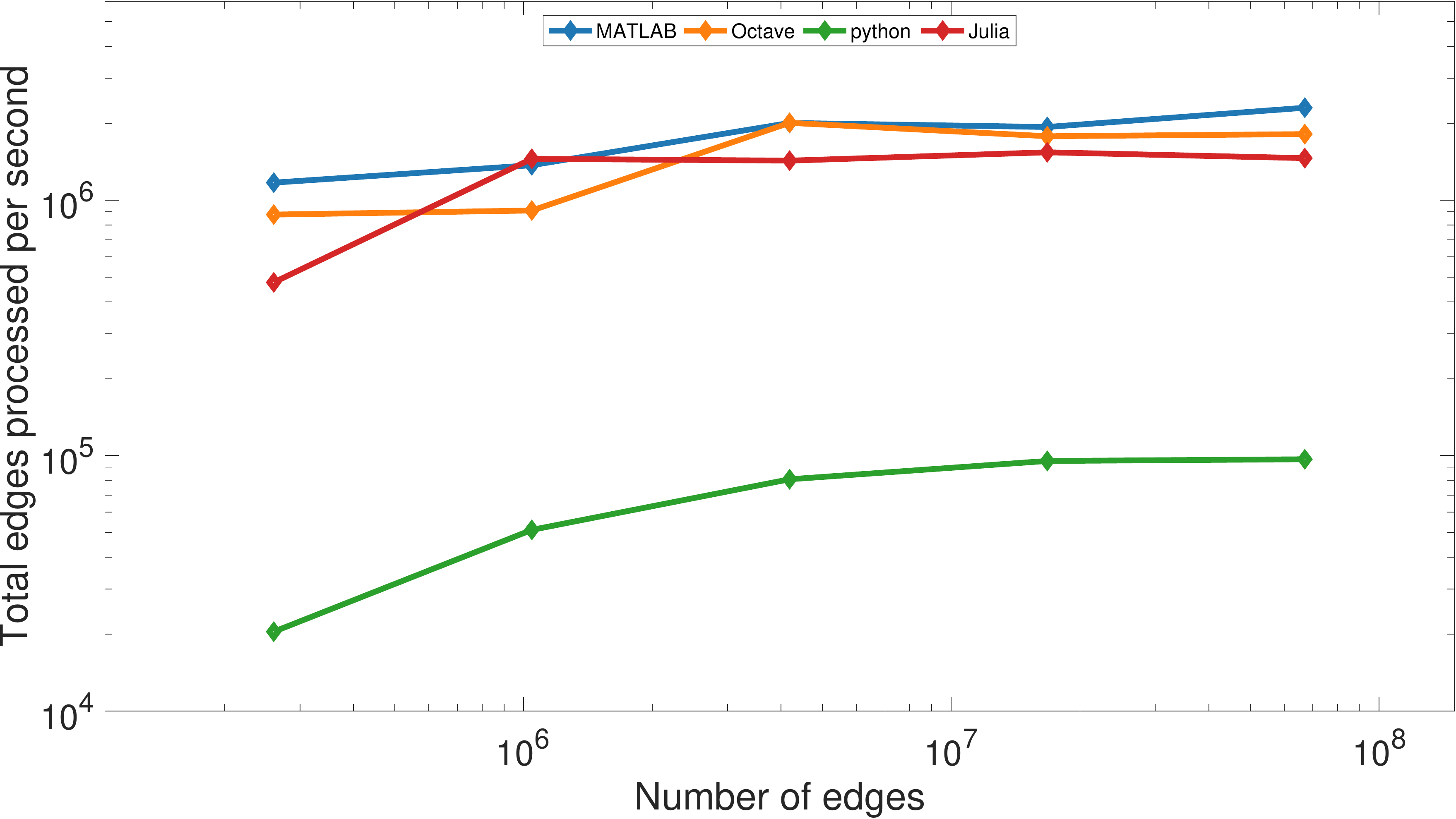}
\caption{k-truss single core performance on synthetic data.}
\label{fig:ktruss_synthetic_edgesort}
\end{figure}

The multi-language implementations of the triangle counting and k-truss algorithms were benchmarked on synthetic graphs. The graphs were generated as $MxM$ images where $M = 2^{n}, n={8,9,10,11,12,13}$. Each pixel in the image was treated as a node in the graph. A pixel was connected to its 8-neighbors by an undirected edge. Table~\ref{table:image_graphs} shows the numbers of edges, nodes and triangles for n = 8 to 13. Examples of two graphs generated in this manner are shown in Figure~\ref{fig:synthetic_graphs}. These graphs were used for testing since the number of nodes, edges and triangles for a given $M$ can be analytically calculated when $M$ is a power of 2. Figure~\ref{fig:triangle_synthetic_edgesort} and Figure~\ref{fig:ktruss_synthetic_edgesort} show the performance of the triangle counting and k-truss implementations in MATLAB, GNU Octave, python and Julia.

\subsubsection{Benchmarking on Real World Data}
\label{sec:snap_data}
\begin{figure}[h]
\centering
\includegraphics[height=11pc]{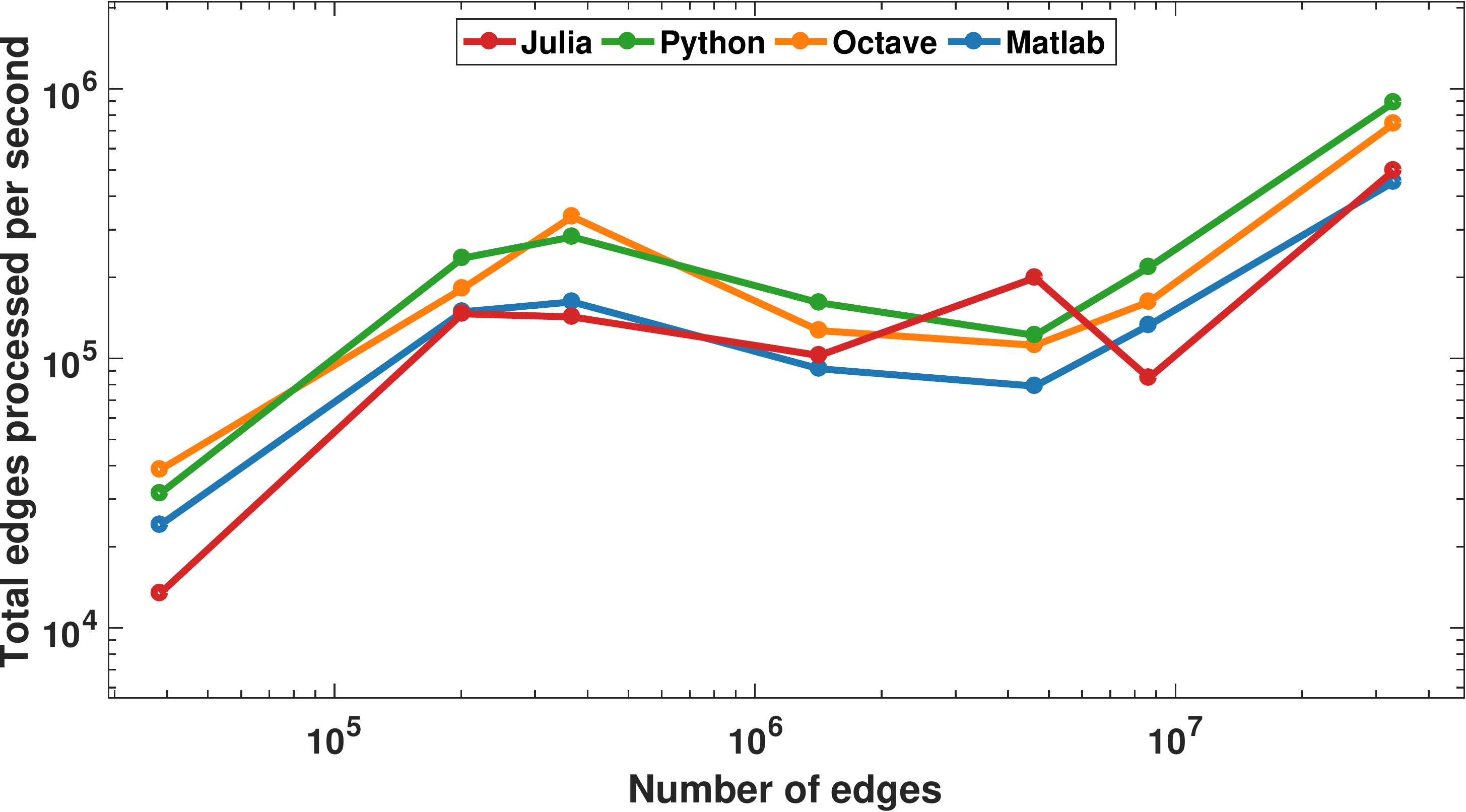}
\caption{Triangle counting single-core performance for a subset of SNAP datasets listed in Table~\ref{data_table}. The average run time of 100 runs for each dataset is shown.}
\label{fig:triangle_perf}
\end{figure}

\begin{figure}[h]
\centering
\includegraphics[height=11pc]{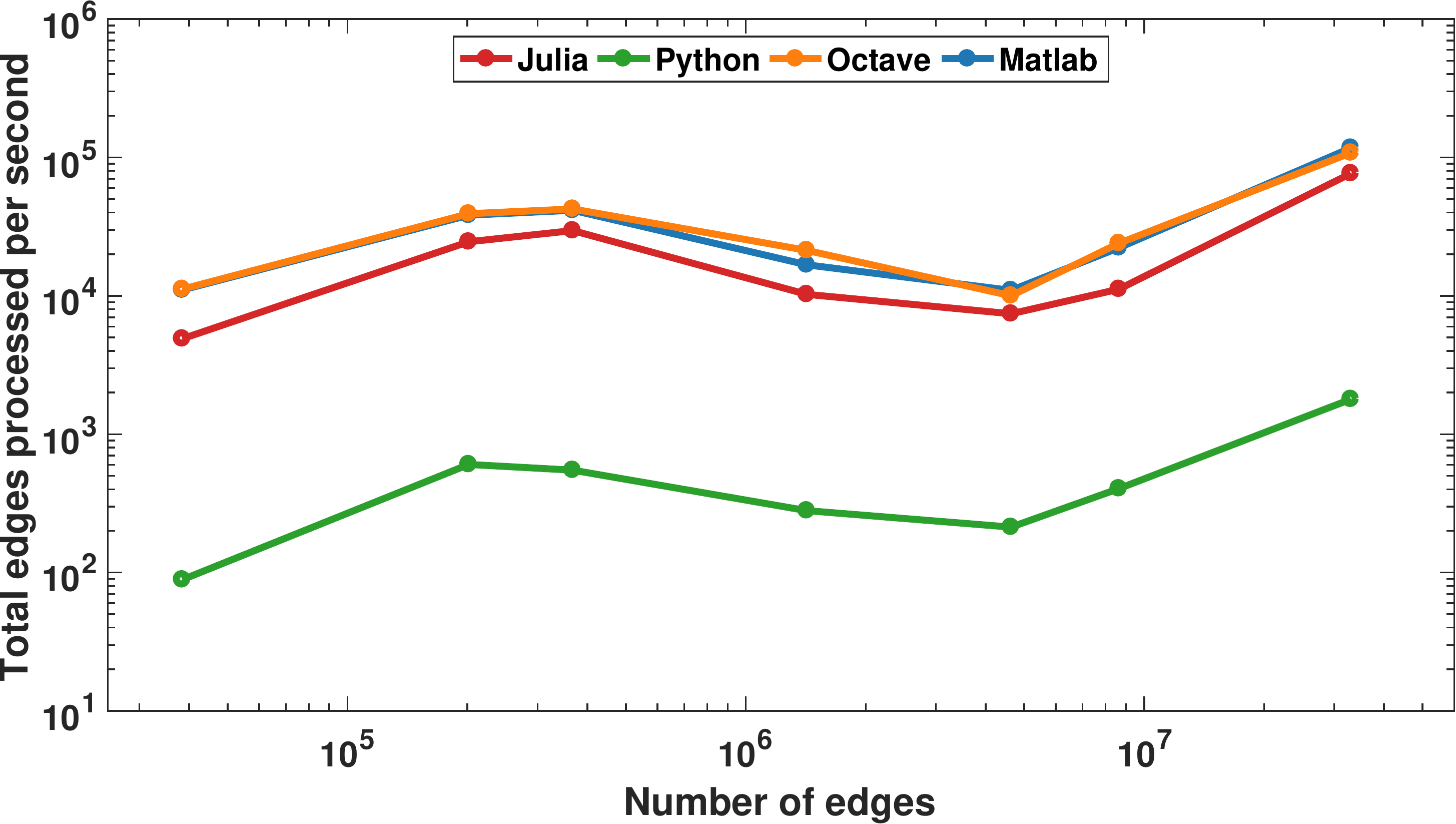}
\caption{k-truss single-core performance for k=3 on a subset of SNAP datasets listed in Table~\ref{data_table}. The average run time of 100 runs for each dataset is shown.}
\label{fig:ktruss_perf}
\end{figure}

Each benchmark was run on a variety of datasets from SNAP~\cite{snapnets}. Graphs
with edge counts ranging from 25,000 to 4.6 million were used for benchmarking
purposes. Figure~\ref{fig:triangle_perf} shows the performance of the triangle
counting benchmark as the ratio of the number of edges in the graph to the total compute time for MATLAB, GNU Octave, python and Julia. Similarly, Figure~\ref{fig:ktruss_perf} shows the performance of
the k-truss implementation in the same languages for $k=3$. The datasets shown in the figures are listed in Table~\ref{data_table}.

\begin{table}
\centering
\def\arraystretch{1.25}
\begin{tabular}{|c|c|c|}
\hline
\textbf{Name} & \textbf{Number of Edges} & \textbf{Number of Triangles}\\
\hline
    cit-HepTh-dates    &     38488  &       1418\\
\hline
    wiki-Vote          &    201524  &    608389\\
\hline
    email-Enron        &    367662  &     727044\\
\hline
    soc-sign-epinions  &   1422420  &    4910076\\
\hline
    flickrEdges        &   4633896  &  107987357\\
\hline
    web-Google         &   8644102  &   13391903\\
\hline
    cit-Patents        &  33037894  &    7515023\\
\hline
\end{tabular}
\caption{Sample SNAP datasets used to generate performance numbers in Figures~\ref{fig:triangle_perf} and \ref{fig:ktruss_perf}}
\label{data_table}
\end{table}

\section{Summary}
The rise of graph analytic systems has created a need for ways to measure and compare
the capabilities of these systems.  Graph analytics present unique scalability
difficulties.  The machine learning, high performance computing, and visual analytics
communities have wrestled with these difficulties for decades and have developed
methodologies for creating challenges to move these communities forward.  The
proposed Subgraph Isomorphism Graph Challenge draws upon prior challenges from
machine learning, high performance computing, and visual analytics to create a graph
challenge that is reflective of many real-world graph analytics processing systems.
The Subgraph Isomorphism Graph Challenge is a mathematically well defined
specification and can be implemented in any programming environment. Subgraph
isomorphism is amenable to both  vertex-centric implementations and array-based
implementations (e.g., using the GraphBLAS.org standard).  The computations are
simple enough that performance predictions can be made based on simple computing
hardware models. Furthermore, since the proposed graph challenge is scalable in both
problem  size and hardware, it can be used to measure and quantitatively compare a
wide range of present day and future systems.  Serial implementations in Python,
Python with Pandas, Matlab, Octave, and Julia have been implemented and their single
threaded performance have been measured. Specifications, data, and software are
publicly available at GraphChallenge.org.

\section*{Acknowledgments}
The authors would like to thank Trung Tran, Tom Salter, David Bader, Jon Berry, Paul
Burkhardt, Justin Brukardt, Chris Clarke, Kris Cook, John Feo,  Peter Kogge, Chris
Long, Jure Leskovec, Richard Murphy, Steve Pritchard, Michael Wolfe, Michael Wright,
and the entire GraphBLAS.org community for their support and helpful suggestions.

\balance

\bibliographystyle{unsrt}
\bibliography{IEEEabrv,references.bib}

\end{document}